\documentclass[12pt,preprint]{aastex}

\slugcomment{Not to appear in Nonlearned J., 45.}

\begin{document}


\title{The Asteroseismology of ZZ Ceti star GD1212}


\author{Lin Guifang\altaffilmark{1,2}, Li Yan\altaffilmark{1,2} and Su Jie\altaffilmark{3}}


\altaffiltext{1}{Yunnan Observatories, Chinese Academy of Sciences, Kunming 650011, China}
\altaffiltext{2}{Key Laboratory for the Structure and Evolution of Celestial Objects, Chinese Academy of Sciences, Kunming 650011, China}
\altaffiltext{3}{Department of Astronomy, Beijing Normal University, Beijing 100875, China}


\begin{abstract}
The ZZ Ceti star GD 1212 was detected to have 19 independent modes from the two-wheel-controlled Kepler Spacecraft in 2014.
By asymptotic analysis, we identify most of pulsation modes. We find out two set of complete triplets, and four sets of doublet which are interpreted as rotation modes with $l=1$. For the other five modes, the four modes $f_{13}$, $f_{15}$, $f_{16}$ and $f_{4}$ are identified as ones with $l=2$; and the mode $f_{7}$ is identified to be the one with $l=1$.  Meanwhile we derive a mean rotation period of $6.65\pm0.21$ h for GD 1212 according to the rotation splitting. Using the method of matching the observed periods to theoretical ones, we obtain the best-fitting model with the four parameters as $M_{\rm{*}}/M_{\rm{\odot}} = 0.775$, $T_{\rm{eff}} = 11400$ K, $\log (M_{\rm{H}}/M_{\rm{*}}) = -5.0$, $\log (M_{\rm{He}}/M_{\rm{*}})=-2.5$ for GD 1212. We find that due to the gradient of C/O abundance in the interior of white dwarf, some modes can not propagate to the stellar interior, which leads to the period spacing of the adjacent modes to become large. This feature is just proven by the observational data from GD 1212. All of these imply that GD 1212 may be evolved from an intermediate mass star.
\end{abstract}


\keywords{asteroseismology-stars:individual(GD1212)-white dwarf}



\section{Introduction}
It's well known that about $98$ percent of stars will become white dwarfs in the end of their evolution \citep{Winget08}. Because there is no thermonuclear burning in the interior of white dwarfs, the evolution of them is basically dominated by cooling. In the cooling curve, pulsating white dwarfs lie in three strips in the H-R diagram \citep{Winget08}: DOVs($75000 \rm{K}\leq T_{\rm{eff}} \leq17000\rm{K}$), DBVs ($22000\rm{K} \leq T_{\rm{eff}}\leq 29000\rm{K}$), and DAVs ($10850\rm{K} \leq T_{\rm{eff}} \leq 12270\rm{K}$). Thereinto about $80$ percent of all pulsating white dwarfs are DAVs or ZZ Ceti stars with the hydrogen atmosphere \citep{Bischoff-Kim2011}. Up to now, there are 148 ZZ Ceti stars \citep{Castanheira2010}.

GD 1212 was confirmed to be a DA-type white dwarf (Holberg et al. 2002; Kawka et al. 2004), and was discovered to pulsate by Gianninas et al. (2006) with the main period of 1060.7s. In recent years, a series of observations by the spectroscopy and multi-color photometry were performed on GD 1212, and its atmospheric parameters(Kawka et al. 2004; Gianninas et al. 2006;Kawka et al. 2007; Subasavage et al. 2009;Sion et al. 2009; Gianninas et al. 2011; Giammichele et al. 2012) can be obtained. Hermes et al. (2014) did the photometric observations on GD 1212 using a two-wheel-controlled Kepler Spacecraft, and showed 19 independent pulsation modes ranging from $828.2 \rm{s} -1220.8 \rm{s}$. In their work, they only gave the discussions about the observed frequencies and various interpretation of the period-spacing, and did not match the pulsation modes with theoretical models.

In this paper, we use the observed data of GD 1212 from the two-wheel-controlled Kepler Spacecraft(Hermes et al. 2014), and try to make the preliminary identification of 19 independent modes in Section 2. In Section 3, we give the calculation of the theoretical model. By comparing the calculated periods of models with the observed ones, we give our best-fitting model.  At last, the summary and discussions are given.

\section{preliminary identification of pulsation modes for GD 1212}

Mode identification is very important in the field of asteroseismology.  In general, the pulsations observed in white dwarfs are non-radial g-mode ones.  The pulsation mode can be characterized by three indices $l, n, m,$ in which $l$ is the spherical harmonic degree, $n$ is the number of the radial nodes, and $m$ is the azimuthal order. If a star is spherical symmetric, $m$ would be degenerate, i.e. modes with the same $l$ and $n$ but different $m$ will have the same frequency. If a star is rotating, rotation will destroy this spherical symmetry and lead to the frequency splitting. In the asymptotic approximation, the formula of frequency splitting due to rotation is as follows (Brickhill 1975)
\begin{equation}
w_{l,n,m} = w_{l,n} + m[1-\frac{1}{l(l+1)}]\Omega
\end{equation}
in which $w_{l,n,m}$ is the frequency of mode with indices $l, n, m$, $\Omega$ is the rotation frequency of white dwarf. It is evident that a frequency with $l=1$ will be splitted as three ones and called a triplet; and a frequency with $l=2$ will be splitted as five ones and called a quintuplet.

According to the asymptotic theory of g-mode, on the other hand, an expression of periods of a mode with $l$ and $n$(Unno et al. 1979; Tassoul 1980) is
\begin{equation}
P_{l,n} \simeq\frac{2\pi^{2}n}{\sqrt{l(l+1)}}(\int_{0}^{R}\frac{N}{r}dr)^{-1}
\end{equation}
in which $N$ is the Brunt-V$\ddot{a}$is$\ddot{a}$l$\ddot{a}$ frequency, and $R$ is the stellar radius.  For a white dwarf, the right integral in the above equation is approximately invariable.  This also means that pulsation periods of different modes with the same $l$ and two adjacent $n$ should have an uniform period spacing. According to Eq.(2), there exists an approximate relation $\triangle \bar{P}(1)/\triangle \bar{P}(2) \sim\sqrt{3}$ between the period spaces of different modes with $l=1$ and with $l=2$.

Using the above asymptotic expressions of g-mode, we reanalyze the 19 independent pulsation modes of GD 1212 from the two-wheel-controlled Kepler Spacecraft(see Table 1 in Hermes et al. 2014 ). Owning to the frequency splitting of rotation, we identify 2 sets of complete triplets and 4 sets of doublet in Table \ref{tbl-1}. For 4 sets of doublet, we assume that the frequency with larger amplitude is the mode with $m = 0$. From Table \ref{tbl-1}, we can obtain that the frequency splitting of modes with $l=1$ due to rotation is about 20 $\rm{\mu Hz}$. According to Eq. (1) and each values of frequency splitting in Table \ref{tbl-1}, we can estimate the rotation period of GD 1212 which are listed in the last column in Table \ref{tbl-1}. In the calculated results, we can notice a rotation period of $8.08$ h which is significantly different from others. This is because of the too small difference of the frequency between the modes of $f_{1}$ and $f_{18}$. According to Table 1 in Hermes et al. (2014), we notice that the error of frequency of $f_{18}$ is 3.54 $\rm{\mu Hz}$. The small difference between $f_{1}$ and $f_{18}$ may be due to the observational error. So we assume that the mode of $f_{18}$ with frequency $857.14$ $\rm{\mu Hz}$ and the other two modes of $f_{14}$ and $f_{1}$ make up of a set of triplet. At last, we estimate the mean rotation period of GD 1212 is $6.65\pm0.21$ h, in which we don't consider the rotation period by the frequency splitting between $f_{1}$ and $f_{18}$.

In 19 pulsation modes, the other five frequencies are unidentified for the absence of frequency splitting. We perform asymptotic analysis for the five pulsation modes. From Table \ref{tbl-2}, we can get the period spacing of modes with $l=1$ is about 37 s.  According to the asymptotic theory of the non-radial g-mode , the period spacing of modes with $l=2$ is about 21 s. For the five unidentified modes, we notice that the period spacing of modes $f_{13}$,$f_{15}$ and $f_{16}$ is about 21 s, which can be identified as ones with $l=2$. Because only one component for these modes has been observed, we hypothesize that they are the ones with $m=0$. The period spacing of the last two modes $f_{4}$ and $f_{7}$ is about 33 s, we cann't firmly identify the two modes as ones with $l=1$ or $l=2$. In the following theoretical calculation, we fit them with $l=1$ or $l=2$.
\begin{table}
\begin{center}
\tablewidth{0pt}
\caption{The possible frequency splitting for $l=1$. ID is the frequency consistent with the one from Table 1 in Hermes et al. (2014). Freq is the frequency in unit of $\rm{\mu Hz}$, d$f$ is the difference of adjacent frequency; Period is the observed period of modes in unit of s; $m$ is the azimuth order, $P_{\rm{rot}}$ is the rotation period of white dwarf  in unit of hours derived by the frequency splitting. \label{tbl-1}}
\begin{tabular}{crrrrr}
\tableline\tableline
ID & Freq($\rm{\mu Hz}$) & d$f$($\rm{\mu Hz}$) & Period(s)  & $m$ & $P_{\rm{rot}}$(h) \\
\tableline
$f_{14}$ & 819.17 &       &  1220.75    &-1 &     \\
        &         & 20.79 &             &   &6.68 \\
$f_{1}$ &839.96   &       &1190.53      &0 &     \\
        &         &17.18  &             &  &8.08  \\
$f_{18}$ &857.14  &       &  1166.67    &+1&      \\
\tableline
$f_{17}$ &888.60  &       &1125.37      &-1 &     \\
        &         &21.85  &             &   &6.36  \\
$f_{2}$ &910.45   &       &1098.36      & 0 &      \\
\tableline
$f_{12}$ &920.71  &       &1086.12      &-1 &   \\
        &         &19.95  &             &   &6.96 \\
$f_{5}$ &940.66  &        &1063.08      &0  &  \\
\tableline
$f_{20}$ &954.02 &        & 1048.19     &-1 &  \\
         &       & 21.29  &             &   &6.52 \\
$f_{9}$ &975.31  &        &1025.31      &0  &   \\
\tableline
$f_{11}$&991.99  &       & 1008.07      &-1 &  \\
        &        & 21.18 &              &   &6.56  \\
$f_{3}$&1013.17  &       & 987.00       & 0 &   \\
\tableline
$f_{8}$& 1166.17 &       &857.51        & -1&  \\
        &        &20.13  &              &   & 6.90 \\
$f_{10}$&1186.30 &       &842.96        &0  &  \\
        &        &21.15  &              &   & 6.57 \\
$f_{6}$ &1207.45 &       &828.19        &+1 &   \\
\tableline
\end{tabular}
\end{center}
\end{table}
\clearpage

\begin{table}
\begin{center}
\tablewidth{0pt}
\caption{The 11 observed frequencies with $m=0$. ID is the frequency consistent with the one from Table 1 in Hermes et al. (2014). Freq is the frequency in unit of $\rm{\mu Hz}$, Period is the observed period of modes in unit of s; d$P$ is the difference of observed adjacent periods; $l$ is the spherical harmonic degree. \label{tbl-2}}
\begin{tabular}{crrrr}
\tableline\tableline
ID & Freq($\rm{\mu Hz}$) &  Period(s)  & d$P$(s)  &  $l$ \\
\tableline
$f_{1}$ &839.96   &  1190.53      &      &1  \\
        &         &               &92.17 &   \\
$f_{2}$ &910.45   &  1098.36      &      & 1  \\
        &         &               &35.28 &     \\
$f_{5}$ &940.66  &   1063.08      &      &1  \\
        &        &                &37.77 &   \\
$f_{9}$ &975.31  &   1025.31      &      &1  \\
        &        &                &38.31 &   \\
$f_{3}$&1013.17  &   987.00      &       &1 \\
       &         &               & 144.04   \\
$f_{10}$&1186.30 &   842.96      &       &1  \\
\tableline
$f_{4}$& 847.17  &  1180.40      &       & ?  \\
       &         &               &  33.28&   \\
$f_{7}$&  871.75 &  1147.12      &       & ? \\
       &         &               & 190.25&       \\
$f_{16}$&1045.08  & 956.87        &       &2  \\
       &         &               & 85.81 &       \\
$f_{15}$&1148.02 &871.06         &       &2 \\
        &         &              & 21.93 &       \\
$f_{13}$&1177.68  & 849.13       &       &2   \\
\tableline
\end{tabular}
\end{center}
\end{table}
\clearpage




\section{model fitting for GD1212}
For research of asteroseismology, it is essential to match the calculated periods of theoretical models to the observed ones in order to find the best-fitting model. After the preliminary identification of all the modes for GD 1212, we have a chance to do the astroseismology study for it.

\subsection{theoretical tools and input physics}
Firstly, we use the suite of open source MESA (Paxton et al. 2011, 2013) to get initial models of white dwarfs. Using the code of MESA, we adopt the metallicity $Z=0.02$ and calculate the evolutions of stars with various masses from zero-age main sequence, to red giant branch, and asymptotic giant branch stages, until they evolve to become hot white dwarfs. The core composition profiles of the hot white dwarfs are the results of the stellar evolution before they become white dwarfs.

Using the initial hot white dwarf models and incorporating the element diffusion (Thoul et al. 1994) into the White Dwarf Evolution Code(WDEC), we calculate the theoretical models of white dwarfs. WDEC was first introduced by Martin Schwarzschild (Schwarzschild \& H$\ddot{a}$rm 1965), and developed by other researchers(e.g. Kutter \& Savedoff 1969; Lamb \& van Horn 1975; Wood 1990; Montgomery 1998; Metcalfe 2001). The code adopts the tables of equation of state (EOS) from Lamb (1974) and Saumon et al. (1995) which are used in the degenerate core and partially ionized envelope, respectively.  And the updated OPAL opacities of Iglesias \& Rogers (1996) are used. In the theoretical calculations, we use the mixing length theory (B\rm{$\ddot{o}$}hm \& Cassinelli 1971) to treat the convection region, and adopt the mixing length parameter $\alpha$ as 0.6(Bergeron et al. 1995) .

In order to calculate the models of DAV white dwarf, WDEC needs to input four parameters: the mass of white dwarf in unit of solar mass $M_{\rm{*}}/M_{\rm{\odot}}$, the effective temperature $T_{\rm{eff}}$, the logarithmic hydrogen mass fraction $\log (M_{\rm{H}}/M_{\rm{*}})$, and the logarithmic helium mass fraction $\log (M_{\rm{He}}/M_{\rm{*}})$.
The corresponding range of the four parameters are:
$0.5 \leqslant M_{\rm{*}}/M_{\rm{\odot}}\leqslant 0.90$ with the step of $\triangle M_{\rm{*}}/M_{\rm{\odot}} = 0.005$, $10850 \rm{K} \leqslant T_{\rm{eff}}\leqslant 12250 \rm{K}$ with $\triangle T_{\rm{eff}} = 50\rm{K}$, $-10\leqslant \log(M_{\rm{H}}/M_{\rm{*}})\leqslant -4$ with $\triangle \log(M_{\rm{H}}/M_{\rm{*}}) = 0.5$, and $-4.0\leqslant \log(M_{\rm{He}}/M_{\rm{*}})\leqslant -2.0$ with $\triangle\log(M_{\rm{He}}/M_{\rm{*}}) = 0.5$. According to the range of the four parameters, we notice that the four parameters cover the most part of the DAV white dwarf. With these grids, and taking the core composition profiles from MESA into account, we use WDEC to calculate all the models of white dwarf. Then we numerically solve the fully equations of linear and adiabatic oscillation in order to find each eigen-mode in theory by scanning in period .

\subsection{the best-fitting model for GD 1212}
In order to find the best-fitting model for GD 1212, we compare the calculated periods for each model with 11 observed ones with previously identified spherical harmonic degree $l$. We introduce the expression of $\chi^{2}$:
\begin{equation}
\chi^{2} = \frac{1}{N} \sum _{i=1}^{N}(P_{\rm{c,i}}-P_{\rm{o,i}})^{2}
\end{equation}
where $P_{\rm{c,i}}$ is the calculated pulsation period from a model; $P_{\rm{o,i}}$ is the observed period; i is an integer which is from 1 to N, and N is the total number of observed periods.

\begin{figure}
\includegraphics[angle=0,scale=1.20]{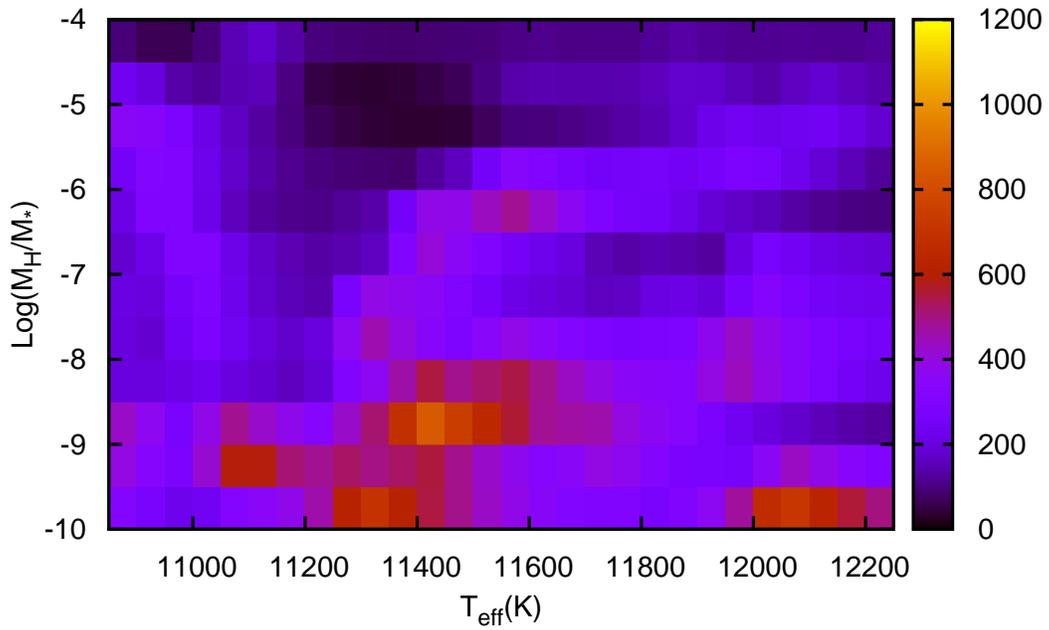}
\caption{The grey-scale map of $\chi^{2}$ for the model with $M_{\rm{*}} = 0.775 M_{\rm{\odot}}$ and $\log (M_{\rm{He}}/M_{\rm{*}}) = -2.5$. The abscissa is the effective temperature $T_{\rm{eff}}$, the ordinate represents the logarithmic hydrogen mass fraction $\log (M_{\rm{H}}/M_{\rm{*}})$.}
\end{figure}

Using the above four-parameter space and the corresponding steps in Section 3.1, we calculate all the models of the white dwarfs and the pulsation periods from every model. Comparing the observed pulsation periods with the theoretical ones, the value of $\chi^{2}$ for each model can be calculated. We choose the model with the minimum $\chi^{2}$ as the best-fitting model, whose parameters are $M_{\rm{*}}/M_{\rm{\odot}} = 0.775$, $T_{\rm{eff}} = 11400$ \rm{K}, $\log (M_{\rm{H}}/M_{\rm{*}}) = -5.0$, $\log (M_{\rm{He}}/M_{\rm{*}})=-2.5$, and the corresponding minimum $\chi^{2} = 4.26$.

Figure 1 gives the grey-scale map of $\chi^{2}$ for the model with $M_{\rm{*}} = 0.775 M_{\rm{\odot}}$ and $\log (M_{\rm{He}}/ M_{\rm{*}}) = -2.5$. The abscissa is the effective temperature $T_{\rm{eff}}$, and the ordinate represents the logarithmic hydrogen mass fraction $\log (M_{\rm{H}}/M_{\rm{*}})$. From Fig. 1, we can see that the space of parameters with lower $\chi^{2}$ lies in the region with $10850 \rm{K} \leq T_{\rm{eff}}\leq 11600 \rm{K}$, and $-6.0\leq\log (M_{\rm{H}}/M_{\rm{*}})\leq-4.0$.

Because the calculated periods of the models can also be influenced by H/He mass fraction, we shorten the steps of the four parameters and reduce their range. The refined range of four parameters are as follows: $0.77 \leqslant M_{\rm{*}}/M_{\rm{\odot}}\leqslant 0.78$ with $\triangle M_{\rm{*}}/M_{\rm{\odot}} = 0.005$, $10850 \rm{K} \leqslant T_{\rm{eff}}\leqslant 11650 \rm{K}$ with $\triangle T_{\rm{eff}} = 50\rm{K}$, $-6\leqslant \log(M_{\rm{H}}/M_{\rm{*}})\leqslant -4$ with $\triangle \log(M_{\rm{H}}/M_{\rm{*}}) = 0.1$, and $-3.0\leqslant \log(M_{\rm{He}}/M_{\rm{*}})\leqslant -2.0$ with $\triangle\log(M_{\rm{He}}/M_{\rm{*}}) = 0.1$. We recalculate all the models and the corresponding pulsation periods. Finally we find that the best-fitting model is the same as the previous one obtained before.

Table \ref{tbl-3} gives the calculated pulsation periods of the best-fitting model. From this table, we notice that the unidentified mode $f_{7}$ with period 1147.12 s can be identified as a mode with $l=1$, $n=27$; the other mode $f_{4}$ with period 1180.4 s can be identified as a mode with $l=2$, $n=48$. It can be noticed that the period difference between theoretical one and observational one for mode $f_{16}$ with period 956.87 s is 3.02 s and for mode $f_{4}$ with period 1180.40 s is 4.08 s, which are comparatively large. However, the observational errors for these two modes are respectively 4.91 s and 4.02 s(Hermes et al. 2014). Our fitting errors are still comparable with the observational errors.

\begin{table}
\begin{center}
\tablewidth{0pt}
\caption{The periods of the best-fitting model. Data of the first and fifth columns are the values of $l$ and $n$ for the calculated periods. $P_{\rm{c}}$ is the calculated pulsation period; $P_{\rm{o}}$ is the observed period. The difference $\mid P_{\rm{c}}-P_{\rm{o}}\mid$ of them is also given. \label{tbl-3}}
\begin{tabular}{cccccccccc}
\tableline
$l = 1$&            &            &                          &$l=2$&            &            &                         \\
$n$    & $P_{\rm{c}}$(s) & $P_{\rm{o}}$(s) & ($\mid P_{\rm{c}}-P_{\rm{o}}\mid$) & $n$ & $P_{\rm{c}}$(s) & $P_{\rm{o}}$(s) & ($\mid P_{\rm{c}}-P_{\rm{o}}\mid$) \\
\tableline
15   & 694.12     &            &                          &34 & 827.75     &            &                \\
16   & 722.13     &            &                          &35 &  849.31      &  849.13  & 0.18           \\
17   & 761.38     &            &                          &36 & 871.62      &   871.06  & 0.56           \\
18   & 795.63     &            &                          &37 & 897.73     &            &                \\
19   &  842.97    &842.96      & 0.01                     &38 &  924.59    &           &     \\
20   & 884.24    &            &                           &39 & 953.85     &   956.87  & 3.02             \\
21   & 919.95    &            &                           &40 & 982.76    &            &              \\
22   & 943.47    &            &                           &41 & 1005.79   &            &             \\
23   &986.57      &987.00      & 0.43                     &42 & 1016.38    &           &  \\
24   &1022.71     &1025.31     & 2.60                     &43 &  1042.68   &           &  \\
25   &1060.40     &1063.08     & 2.68                     &44 &  1069.46  &            &         \\
26   &1099.05     &1098.36     & 0.69                     &45 &   1096.79  &           &   \\
27   & 1148.18   &  1147.12   & 1.06                     &46 &  1123.37   &            &            \\
28   &1192.78     &1190.53     & 2.25                     &47 &  1150.43   &            &          \\
29   &1232.93     &            &                          &48 & 1176.32    &  1180.40   & 4.08     \\
30   &1256.67     &            &                          &49 & 1200.64    &            &          \\
\tableline
\end{tabular}
\end{center}
\end{table}

\subsection{Constraints from the theoretical models}
Figure 2 gives the period spacing of calculated consecutive modes with $l=1$ and $l=2$. The upper panel is the one with $l=1$ and the lower panel is the one with $l=2$. The hollow circles represent the period spacing of theoretical modes; and the filled circles denote the one of observed periods with adjacent modes. For the panel with $l=1$, the number of radial nodes $n$ is from 19 to 31, and for the other with $l=2$, $n$ is from 33 to 53. In this figure, we notice that results of theoretical models are in good agreement with the observations. Furthermore, we notice that the first three observed period spacings are around 37 s, then the next two observed period spacings suddenly increase to 48.76 s and 43.41 s. Such large increase in the period spacing usually implies changes of propagation region for different pulsation modes. We will discuss this feature in the following.

\begin{figure}
\centering
\begin{minipage}[c]{0.8\textwidth}
\centering
\includegraphics[angle=0,scale=1.10]{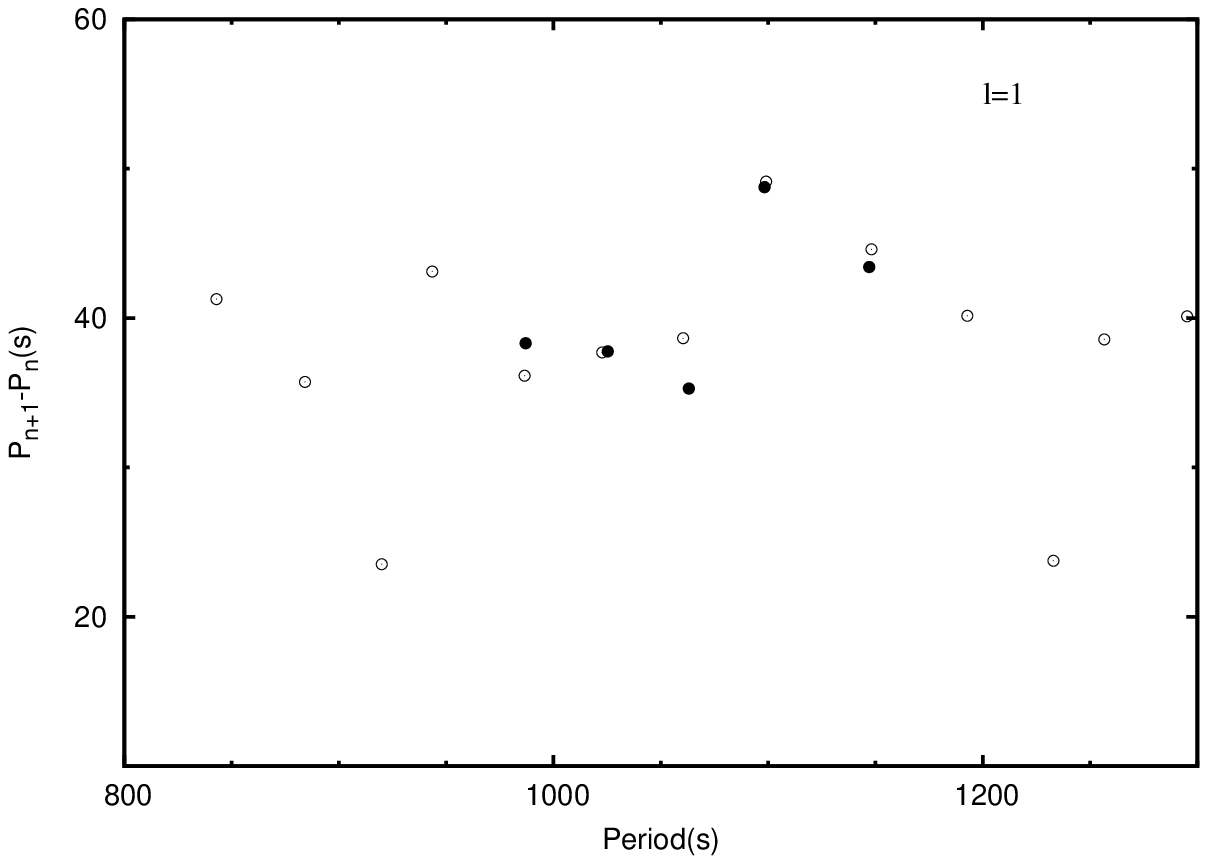}
\end{minipage}
\begin{minipage}[c]{0.8\textwidth}
\centering
\includegraphics[angle=0,scale=1.10]{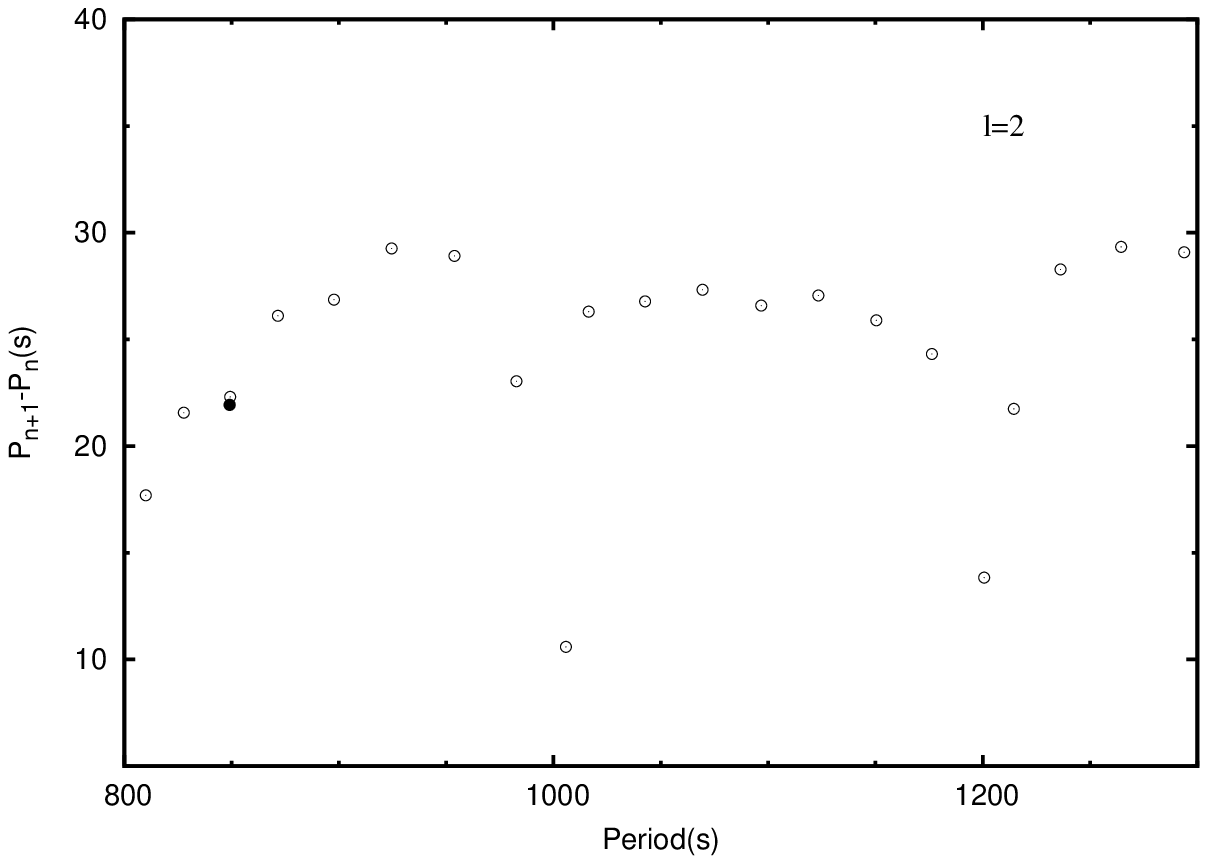}
\end{minipage}
\caption{The period spacing of consecutive modes with $l=1$ (upper) and $l=2$ (lower). The hollow circles represent the period spacing of theoretical modes and the filled circles denote the one of observed periods with adjacent modes.}
\end{figure}

Figure 3 shows the composition profiles in the best-fitting model with $M_{\rm{*}} = 0.775 M_{\rm{\odot}}$. We can see that there are 2 transition zones in the interior of the white dwarf. One is the transition zone of H and He which lies in the position of $\log (1-M_{\rm{r}}/M_{\rm{*}}) \sim -5$. The other is the one of He and C/O which is located at the position of $\log (1-M_{\rm{r}}/M_{\rm{*}}) \sim -2.5$. It should be noticed that the core of the white dwarf is the result of nuclear burning and convective mixing in the stellar evolution. Before the star becomes a white dwarf, a convection core lies in the central part of the star, which makes the elements completely mix in the core. Outside the convection core, there is a region where a gradient of C/O abundance forms.

\begin{figure}
\includegraphics[angle=0,scale=1.20]{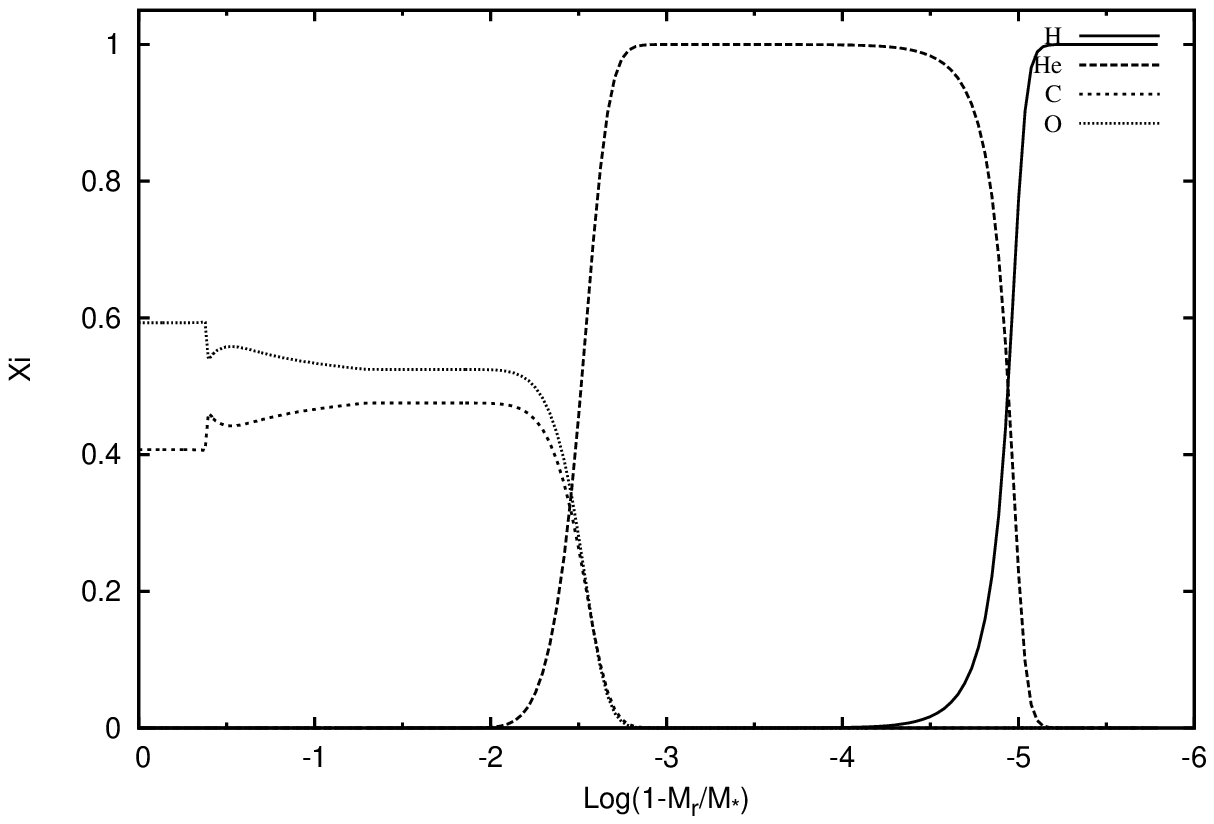}
\caption{The composition profiles in the best-fitting model with $M_{*} = 0.775 M_{\odot}$.}
\end{figure}

In order to explain the feature that the period spacing of consecutive modes with $l=1$ and $n=26$ suddenly becomes larger in Fig. 2, we calculate the so-called modal mass $M_{\rm{mode}}$, whose expression is:
\begin{equation}
M_{\rm{mode}} =4\pi\int_{0}^{R}[\mid \xi_{r}\mid^{2}+l(l+1)\mid\xi_{h}\mid^{2}]\rho_{0}r^{2}dr
\end{equation}
in which $\xi_{r}$ is the radial displacement, $\xi_{h}$ is the horizontal displacement, $\rho_{0}$ is the stellar density, and $r$ is the radius.

The distributions of the so-call modal mass for some modes with $l = 1$ are shown in Fig. 4. The abscissa is the nondimensional quantity $\log (1-M_{\rm{r}}/M_{\rm{*}})$. The ordinate represents a normalized integrant of the so-call modal mass. The panels from top to bottom are the ones for modes with $n$ from 23 to 30. The scales in the ordinate and abscissa are the same for all panels. From this figure, we can notice that most of pulsation energy for all modes are distributed in two transition zones. One of them lies in the position of $\log (1-M_{\rm{r}}/M_{\rm{*}}) \sim -5$, which is the transition zone of H and He. The other is around $\log (1-M_{\rm{r}}/M_{\rm{*}}) \sim -2.5$, which is the one of He and C/O. We also find that the pulsation energy in the interior of C/O core for modes with $n=26$ and $n=27$ is very small compared with others. Meanwhile their period spacings with respect to the adjacent mode increase suddenly. This feature is evident very clearly also in the observed data (see Fig. 2). On the other hand, the pulsation energy in the core for the mode with $n=29$ is very large, and the corresponding period spacing is very small. All of these can be explained by Eq. (2). When the pulsation energy can be propagated into the stellar interior, the range of the integral in Eq. (2) must be extended to the stellar core. The integral part becomes very large, which leads to a small period spacing, and vice versa.

Comparing Fig. 3 with Fig. 4, we notice that the region with such a gradient of C/O abundance lies in the position of $\log (1-M_{\rm{r}}/M_{\rm{*}}) \sim -0.5$. It separates the stellar core into two parts: the inner core has a lower C/O ratio and the outer core has a higher C/O ratio. Due to the gradient of C/O abundance, some modes can not propagate to the stellar interior. This leads to the period spacing for these modes to become large. This feature is just proven by the observational data from GD 1212. It means that the gradient of C/O abundance should be exist in the interior of GD 1212. For GD 1212, the mass of the best-fitting model is $M_{*}/M_{\odot}=0.775$. In the central He burning phase, there is an convection core where the elements are to be mixed completely. The profiles of C/O elements outside the convection core will be formed by the shell burning of He element in the phase of AGB. All of these imply that GD 1212 may be evolved from an intermediate mass star.

\begin{figure}
\includegraphics[angle=0,scale=1.1]{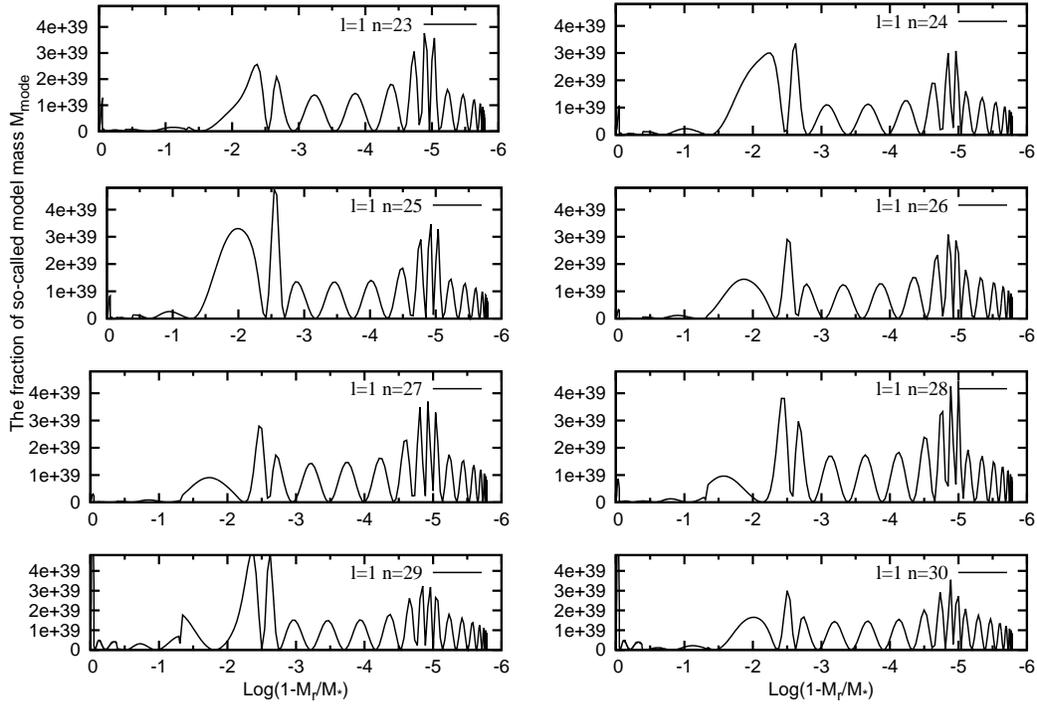}
\caption{The distribution of the fraction of the so-call modal mass. The abscissa is the nondimensional quantity $\log (1-M_{\rm{r}}/M_{\rm{*}})$. The ordinate represents a fraction of the so-call modal mass. The panels from top to bottom are the ones for modes with $l=1$ and $n$ from 23 to 30.}
\end{figure}





\section{Summary and discuss }
Asteroseismology provides a unique window to detect the internal structure of the pulsating stars by matching the observed periods to theoretical periods generated by adiabatic pulsation.

In this paper, by asymptotic analysis, we identify most of pulsation modes of GD 1212 observed by the two-wheel-controlled Kepler Spacecraft (Hermes et al. 2014). Because of the rotation splitting, we identify two set of complete triplets and four sets of doublet. We interpret them as rotation modes with $l=1$ owing to the resemble frequency splitting. Thus we estimate a mean rotation period is $6.65\pm0.21$ h for GD 1212. Using asymptotic analysis of non-radial g-mode, we identify the three modes $f_{13}$, $f_{15}$ and $f_{16}$ as ones with $l=2$. The two modes $f_{4}$ and $f_{7}$ are identified as ones with $l=2$ and $l=1$, respectively, by matching the observed periods with theoretical ones. By the asteroseismology study, four parameters of the best-fitting model for GD 1212 can be obtained as $M_{\rm{*}}/M_{\rm{\odot}} = 0.775$, $T_{\rm{eff}} = 11400$ K, $\log (M_{\rm{H}}/M_{\rm{*}}) = -5.0$, $\log (M_{\rm{He}}/M_{\rm{*}})=-2.5$. Simultaneously, we find that due to the gradient of chemical element abundance outside the convection zone, some modes can not propagate to the stellar interior. This leads to the period spacing of the adjacent modes to become large, which is just proven by the observational data from GD 1212. This means that the gradient of C/O abundance should exist. And ZZ Ceti star GD 1212 may be evolved from an intermediate mass star.

\begin{deluxetable}{cccccc}
\tablewidth{0pt}
\tablecaption{The values of the parameters from the various observations.\label{tbl-4}}
\tablehead{\colhead{$T_{\rm eff}$(K)}
          &  \colhead{$\log$ $g$}
          &     \colhead{$M/M_{\odot}$}
          & \colhead{$\log M_{\rm H}/M_{*}$}
          &\colhead{$\log M_{\rm He}/M_{*}$}
          &\colhead{Ref} }
\startdata
10960$\pm$75 &8.2$\pm$0.1    &  0.73$\pm$0.06  &    &    & 1\\
11040$\pm$132&8.11$\pm$0.038 & 0.67 $\pm$0.02  &    &    & 2\\
11010$\pm$210&8.05$\pm$0.15  &0.63$\pm$0.09    &    &    &3\\
11000$\pm$300&8.25$\pm$0.04  &0.76$\pm$0.02  &      &    &4\\
11270$\pm$165&8.18$\pm$0.05  &0.71$\pm$0.03    &    &     &5\\
10938$\pm$317&8.25$\pm$0.03  &0.76$\pm$0.02    &    &     &6\\
11400        & 8.37          &0.775            &-5.0&-2.5 &7\\
\enddata
\tablerefs{
(1)Kawka et a. 2004;(2)Gianninas et al. 2006;(3)Kawka et al. 2007;(4)Subasavage et al. 2009;(5)Gianninas et al. 2011;(6)Giammichele et al. 2012;(7)our model}
\end{deluxetable}

In order to compare the theoretical parameters with ones obtained from observation for GD 1212, we collect the values of the parameters from the various observations. In Table \ref{tbl-4}, we list the values of parameters: the effective temperature $T_{\rm{eff}}$, $\log g$, and the mass $M_{\rm{*}}$ obtained from the spectrum and multi-color photometry. The last row is our calculated results for GD 1212. From this table, our result is very close to the ones from Subasavage et al. (2009) and Giammichele et al. (2012).In future, further observations of GD 1212 are needed. This will help us to find more independent modes and then further constrain the model of GD 1212.



\acknowledgments

This work is supported by the Knowledge Innovation Key Programme of the University of Chinese Academy of Sciences under Grant No. KJCX2-YW-T24 and the Foundation of XDB09010202. We are very grateful to Yanhui Chen and Tao Wu for their kindly discussion and suggestions.

\clearpage

\end{document}